\documentclass[prb,aps,epsf,eqsecnum,twocolumn,showpacs,ltxgrid,epsfig,psfrag]{revtex4}
\usepackage{amsmath}
\usepackage{epsfig}          
\begin{document}   

\title{Linear response of a grafted semiflexible polymer to a uniform force field}
\author{Panayotis Benetatos$^1$ and Erwin Frey$^{1,2}$}
\affiliation{$^1$Hahn-Meitner-Institut, Abteilung Theoretische Physik,
Glienicker Str. 100, D-14109 Berlin, Germany\\
$^2$Fachbereich Physik, Freie Universit\"at Berlin, Arnimallee 14, D-14195 Berlin, Germany}


\date{\today}

\begin{abstract}

We use the worm-like chain model to analytically calculate the linear response of a grafted semiflexible polymer to a uniform force field. The result is a function of the bending stiffness, the temperature, the total contour length, and the orientation of the field with respect to that of the grafted end. We also study the linear response of a worm-like chain with a periodic alternating sequence of positive and negative charges. This can be considered as a model for a polyampholyte with intrinsic bending siffness and negligible intramolecular interactions. We show how the finite intrinsic persistence length affects the linear response to the external field.

\end{abstract}

\maketitle

\section{Introduction}

Semiflexible polymers are macromolecules with a bending stiffness intermediate between that of an absolutely flexible (Gaussian) chain and a rigid rod. A measure of the directedness of a polymer is its persistence length, which is proportional to the bending stiffness. Semiflexible polymers have a presistence length of the order of the total contour length. They have been the subject of extensive theoretical and experimental study in recent years primarily because many biologically important macromolecules fall in this class. The structural elements of the cytoskeleton (actin filaments, microtubules, intermediate filaments) and double-helix DNA exhibit elastic behavior dominated by their bending stiffness \cite{JH,PhN}.

Single-molecule experiments have allowed the study of the influence of external forces or force fields on the conformational properties of semiflexible polymers\cite{Bust,Chu,Lad,Fer}. Their theoretical study is considerably more challenging compared to that of flexible chains and one of the main reasons is the lack of scaling properties except for limiting cases (e.g., weakly bending polymers). In the worm-like chain model, pulling a semiflexible polymer at its ends is formally analogous to the Stark effect of a quantum rotator \cite{Fixman}. This analogy has led to semi-analytical solutions of the force-extension problem. The response to a force field is even more complicated as the quantum analogy involves a non-local (in imaginary time) interaction. Further complications arise from intramolecular interactions between different polymer segments which may become important in realistic experimental situations where a polymer is subject to a hydrodynamic flow or it is charged and placed in an electric field. 

Marko and Siggia \cite{MS} have derived an approximate analytical field-extension relation for a semiflexible polymer which has one end fixed and is free to rotate about it. Their result for strong fields appears to be in good agreement with an experiment done with DNA in an external electric field \cite{Udo}. Lamura {\it et al.} \cite{Lamura} have calculated conformational properties of a grafted semiflexible polymer in a uniform force field in two dimensions using recursion relations in the weakly bending approximation which is valid at strong fields and/or large persistence lengths. We should mention that the standard Gaussian polymer model is inadequate to describe stretching in strong fields because it is infinitely extensible \cite{Mans,Gav}. In contrast, the worm-like chain is  characterized by a fixed total contour length constraint and this problem is avoided. 

Charged polymers with two types of charge (positive and negative) along their backbone are called polyampholytes. They have attracted a lot of attention because of their inherent theoretical interest and their relevance to the protein folding problem \cite{KanKar}. There have been studies of their response to an electric field \cite{Schiess1,Schiess2,Wink,Rub} but most of the previous works deal with their conformational properties due to the intramolecular interactions in the absence of an external field. They all consider charge distributions on backbones described as flexible or freely-jointed chains and do not take into account the possibility of an internal bending stiffness.  

In the present paper we calculate the linear response of a grafted worm-like chain to a uniform electric field for arbitrary bending stiffness and field orientation. In Sec. II we describe the model. In Sec. III we calculate the linear response of the orientation and the extension of a uniformly charged worm-like chain and compare the results with the response to a force applied at the free end. In Sec. IV we calculate the linear response of the extension of a periodic alternating polyampholyte with internal bending stiffness.

\section{The model}

The worm-like chain  is a fluctuating, continuous, locally inextensible line with a fixed total contour length $L$. In the absence of any external force, its effective free energy functional depends only on the bending (curvature) and is given by 
\begin{equation}
\label{WLCHamfree}
{\cal H}_0 [\{{\bf r(s)}\}]= \frac{\kappa}{2}\int_0^L ds \Big [ \frac{\partial {\bf t}(s)}{\partial s} \Big ]^2 \;,
\end{equation}
where ${\bf t}(s)={\partial {\bf r}(s)}/{\partial s}$ is the unit tangent vector of the curve ${\bf r}(s)$ at arc length $s$, and $\kappa$ is the bending stiffness \cite{Saito}. The correlation length of the unit tangent vector along the polymer contour is the persistence length, $L_p$, which is related to the bending stiffness via $L_p=2\kappa/[k_BT(d-1)]$, where $d$ is the dimensionality of the embedding space. Throughout this paper, we consider a worm-like chain which is grafted at $s=0$, that is, both the position of this point {\it and} the orientation of the related tangent vector are kept fixed.

The interaction with a uniform external field ${\bf E}$ is expressed  by adding to ${\cal H}_0$ the term
\begin{eqnarray}
\label{WLCHamint}
& &{\cal H}_I [\{{\bf r(s)}\}]= -{\bf E}\cdot \int_0^L ds \lambda(s)[ {\bf r}(s) - {\bf r}(0)]\nonumber\\
& & =-{\bf E}\cdot \int_0^L ds \lambda(s) \int_0^s ds'{\bf t }(s')\;,
\end{eqnarray}
where $\lambda(s)$ is a phenomenological linear charge density. This simple model neglects the intramolecular interaction between different segments of the same polymer and also the steric self-avoidance. The latter is expected to be negligible for $L_p>L$. For polymers with a large aspect ratio, like double-helix DNA or F-Actin, self-avoidance effects become important only for $L\gg L_p$.

If $\theta(s)$ is the angle between the tangent vector ${\bf t}(s)$ and the field ${\bf E}$, we can write
\begin{equation}
\label{WLCH}
{\cal H}= {\cal H}_0 - E \int_0^L ds \lambda(s) \int_0^s ds' \cos\theta(s') \;.
\end{equation}

The orientational probability distribution of a worm-like chain having a conformation with a tangent vector ${\bf t}(0)={\bf t}_0$ at one end and a tangent vector ${\bf t}(L)={\bf t}_L$ at the other is given by a path integral which is formally analogous to the density matrix element, in the angle representation, of a quantum rigid rotator \cite{Saito}. We denote it by $G({\bf t}_L, L|{\bf t}_0, 0)$. It can be calculated analytically and it has a spectral representation in terms of spherical harmonics. In $3$ dimensions, we can integrate out the azimuthal angle to obtain $G[\theta(s), s|\theta(s'), s']$. Thus we are able to calculate orientational correlations using 
\begin{eqnarray}
\label{prop}
& &\langle \cos\theta(s_n)...\cos\theta(s_1)\rangle_0  \nonumber\\
& &=\int d\theta(L)\sin\theta(L)\int d\theta(s_n)\sin\theta(s_n)...\int d\theta(s_1)\sin\theta(s_1)\nonumber\\
& &\times G[\theta(L), L|\theta(s_n), s_n]\cos\theta(s_n)G[\theta(s_n), s_n|\theta(s_{n-1}), s_{n-1}]\nonumber\\
& &\times ...\cos\theta(s_1)G[\theta(s_1), s_1|\theta(0), 0]\;,
\end{eqnarray}
where $L>s_n>s_{n-1}>...>s_1>0\;$ \cite{Saito}.

\section{Linear response of a uniformly charged semiflexible polymer}

We define the change in the orientation of the polymer due to the applied force field by 
\begin{equation}
\label{lrdef}
\delta\langle\cos\theta(s)\rangle \equiv \langle\cos\theta(s)\rangle_E-\langle\cos\theta(s)\rangle_0\;,
\end{equation}
where the first average in the rhs of the equation is taken over the Boltzmann weight associated with the energy given in Eq. (\ref{WLCH}) and the second average is taken over the Boltzmann weight associated with the energy given in Eq. (\ref{WLCHamfree}).
To lowest order in $E$, 
\begin{eqnarray}
\label{lr}
& &\delta\langle\cos\theta(s)\rangle=\frac{E\lambda}{k_B T}\Big\{\int_0^Lds'\int_0^{s'}ds''\langle\cos\theta(s)\cos\theta(s'')\rangle_0\nonumber\\
& &-\langle\cos\theta(s)\rangle_0\int_0^Lds'\int_0^{s'}ds''\langle\cos\theta(s'')\rangle_0\Big\}\;.
\end{eqnarray}
In this Section, we assume that the linear charge density $\lambda$ is constant along the polymer contour. Using the prescription of Eq. (\ref{prop}), one obtains
\begin{equation}
\label{aux1}
\langle\cos\theta(s)\rangle_0=\cos\theta(0)\exp(-s/L_p)
\end{equation}
and
\begin{eqnarray}
\label{aux2}
& &\langle\cos\theta(s)\cos\theta(s')\rangle_0=\exp[-(s-s')/L_p]\nonumber\\
& &\times\Big\{\frac{1}{3}+\exp(-3s'/L_p)\big[\cos^2\theta(0)-\frac{1}{3}\big]\Big\}\;,
\end{eqnarray}
where $s>s'\;.$ Substituting Eqs. (\ref{aux1}) and (\ref{aux2}) into  Eq. (\ref{lr}) and breaking the domain of integration of the first double integral in the rhs of the equation into parts with a well-defined arc-length ordering, we obtain the final result:
\begin{eqnarray}
\label{res1}
& &\delta\langle\cos\theta(s)\rangle\nonumber\\
& &=\frac{E\lambda L_p^2}{k_BT}\Big\{-\frac{2}{3}\frac{s}{L_p}+\frac{2}{3}\frac{L}{L_p}-\frac{2}{3}\frac{L}{L_p}\exp(-s/L_p)\nonumber\\
& &+\frac{1}{3}\exp[(s-L)/L_p]-\frac{1}{3}\exp[-(s+L)/L_p]\nonumber\\
& &+\big(\cos^2\theta(0)-\frac{1}{3}\big)\big[\frac{1}{2}\frac{L}{L_p}\exp(-3s/L_P)\nonumber\\
& &+\exp[-(2s+L)/L_p]-\frac{3}{4}\exp(-3s/L_p)\nonumber\\
& &+\frac{3}{4}\exp(-s/L_p)-\exp[-(s+L)/L_p]\nonumber\\
& &-\frac{1}{2}\frac{L}{L_p}\exp(-s/L_p)-\frac{1}{2}\frac{s}{L_p}\exp(-3s/L_p)\big]\Big\}\;.
\end{eqnarray}

It is intstructive  to consider the flexible and the weakly bending limits of this result. In the flexible limit  where $L\gg L_p$, the anisotropy associated with the $\theta(0)$-dependence of the response drops out and we get 
\begin{equation}
\label{flex1}
\delta\langle\cos\theta(L)\rangle=\frac{E\lambda L_p^2}{3k_BT}+O\Big(\big(\frac{L}{L_p}\big)\exp\big(-\frac{L}{L_p}\big)\Big)\;.
\end{equation}
The leading term is precisely the response of a freely rotating dipole moment $\mu=\lambda L_p^2$ to a weak field ${\bf E}\,.$ The average polarization of such a system is $\langle\mu\cos\theta\rangle=(1/2) \int d \theta \sin \theta \mu \cos \theta \exp[-\mu E \cos \theta/(k_BT)]$. 
On the other hand, in the weakly bending limit where $L \ll L_p$, the response strongly depends on $\theta(0)$. For $\theta(0)\neq (0, \pi)$, 
\begin{equation}
\label{stiff1a}
\delta\langle\cos\theta(L)\rangle=\frac{E\lambda L^3}{6 \kappa}\sin^2\theta(0)+O\Big(\big(\frac{L}{L_p}\big)^4\Big)\;.
\end{equation}
For a purely bending field ($\theta(0)=\pi/2$), we recover the mechanical bending of a rod with siffness $\kappa$ which is known from the classical theory of elasticity \cite{LL}. For an elongational (or compressional) field with $\theta(0)=(0, \pi)$, 
\begin{equation}
\label{stiff1b}
\delta\langle\cos\theta(L)\rangle=\frac{E\lambda L^4 k_BT}{12 \kappa^2}+O\Big(\big(\frac{L}{L_p}\big)^5\Big)\;.
\end{equation}
We note that in this case the linear response becomes proportional to the temperature and it vanishes at zero temperature. This vanishing is an indication of the existence of a {\it finite} critical field above which the compressional deformation becomes unstable as predicted by the classical theory of elasticity of rods (buckling to a field) \cite{LL}.


\begin{figure}
\begin{center}
\label{deltaR}
\leavevmode
\hbox{%
\epsfxsize=2.5in
\epsffile{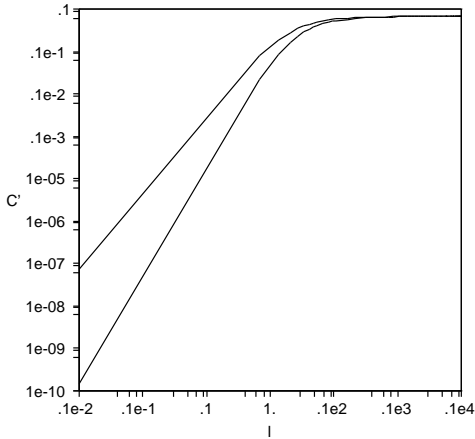}}
\end{center}
\caption{Rescaled linear response coefficient for the extension, $C'\equiv \delta R_{\parallel} k_B T/(E\lambda L_p L^2)$, as a function of the chain stiffness, $l\equiv L/L_p$, for $\theta_0=\pi/2$ (upper curve) and $\theta_0=0$ (lower curve).}
\end{figure}


An experimentally more accessible quantity is the response of the end-to-end vector in the direction of the field:
\begin{equation}
\label{lrRdef}
\delta R_{\parallel}\equiv \big(\langle{\bf R}\rangle_E-\langle{\bf R}\rangle_0\big)\cdot \frac{\bf E}{|{\bf E}|}\;,
\end{equation}
where
\begin{equation}
\label{R}
{\bf R}=\int_0^Lds {\bf t}(s)\;.
\end{equation}
We easily obtain it by integrating Eq. (\ref{res1}) over the polymer contour:
\begin{eqnarray}
\label{lr2}
& &\delta R_{\parallel}=\frac{E\lambda L_p^3}{k_BT}\Big\{-\frac{2}{3}\big[\exp(-L/L_p)+\frac{L}{L_p}\nonumber\\
& &-\frac{L}{L_p}\exp(-L/L_p)\big]+\frac{1}{3}\big[\big(\frac{L}{L_p}\big)^2+\exp(-2L/L_p)\nonumber\\
& &+36\big]+ \big(\cos^2\theta(0)-\frac{1}{3}\big)\big[-\frac{1}{3}\frac{L}{L_p}\nonumber\\
& &-\frac{5}{4}\exp(-L/L_p)+\frac{4}{9}- \frac{7}{36}\exp(-3L/L_p)\nonumber\\
& &+\exp(-2L/L_p)+\frac{1}{2}\frac{L}{L_p}\exp(-L/L_p)\big]\Big\}\;.
\end{eqnarray}

As with the response of the orientation, we gain insight by considering the limiting cases of flexible and weakly bending polymers. In the flexible limit ($L\gg L_p$),\begin{equation} 
\label{flex2}
\delta R_{\parallel}=\frac{E\lambda L_p L^2}{3 k_BT}+O\big(\frac{L}{L_p}\big)\;.
\end{equation}
The leading term is the result that one obtains for the response to a uniform field of a Gaussian chain with Kuhn length equal to $2L_p\,.$ Obviously, this result cannot be valid in the strong field regime because the $\sim L^2$ scaling would become incompatible with the local inextensibility of the polymer. Indeed, the response of a worm-like or a freely-jointed chain  to a strong field has different scaling behavior and is free from this pathology as it is shown in Refs. \cite{MS}, \cite{Mans}, and \cite{Gav}. For weak fields, however, there is no inconsistency between Eq. (\ref{flex2}) and the constraint of fixed contour length. In the weakly bending limit ($L\ll L_p$), 
\begin{equation}
\label{stiff2a}
\delta R_{\parallel}=\frac{E\lambda L^4}{8\kappa}\sin^2\theta(0)+O\Big(\big(\frac{L}{L_p}\big)^5\Big)
\end{equation}
for $\theta(0)\neq (0, \pi)$, and
\begin{equation}
\label{stiff2b}
\delta R_{\parallel}=\frac{E\lambda k_BT L^5}{20 \kappa^2}+O\Big(\big(\frac{L}{L_p}\big)^6\Big)
\end{equation}
for $\theta(0)= (0, \pi)$. As before, the bending response in the weakly bending  limit reproduces the result from the classical elasticity of rods \cite{LL} and the temperature dependence in Eq. (\ref{stiff2b}) is a sign of its entropic origin. The crossover between the two scaling limits as the chain stiffness varies is illustrated in Fig. 1.


\begin{figure}
\begin{center}
\label{qeff}
\leavevmode
\hbox{%
\epsfxsize=2.5in
\epsffile{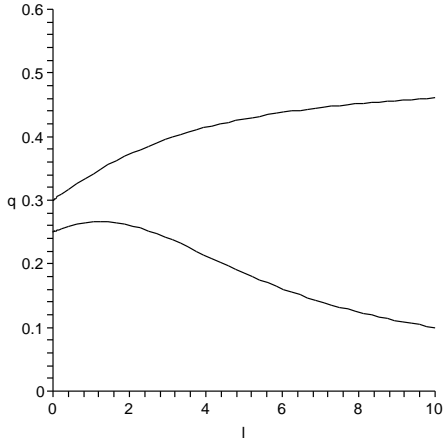}}
\end{center}
\caption{ Effective point charge for the response of the extension (upper curve) and for the response of the orientation (lower curve) as a function of $l\equiv L/L_p$ for $\theta_0=0$.}
\end{figure}


In order to compare the linear response to a field to the linear response to a force exerted at the free end, it is instructive to define an ``effective point charge.'' It is a point charge placed at the free end which, subject to the field ${\bf E}$, would yield the same response as that due to the total charge $\lambda L$ which is uniformly distributed along the polymer contour. In Fig. 2, we plot the ``effective point charge'' (in units of $\lambda L$) as a function of $L/L_p$ for the response of the extension, $ \delta R_{\parallel}$, (upper curve) and for that of the orientation, $\delta \langle \cos \theta(L) \rangle$, (lower curve). The equations giving the response to a force applied at the free end of a grafted worm-like chain  are cited in Appendix A. We note that the effective charge for the extension tends to $(1/2)\lambda L$ in the flexible limit which is just the average charge along the polymer contour. In the weakly bending limit, it decreases to $(3/10)\lambda L$ for $\theta(0)=(0, \pi)$ and to $(3/8)\lambda L$ for any other orientation.

\section{Alternating polyampholytes}

In this Section, we extend our study of a charged worm-like chain in an external electric field to a grafted semiflexible polymer with a periodic alternating charge sequence along its contour. We model such a sequence by a sinusoidal linear charge density:
\begin{equation}
\label{chd}
\lambda(s)=\lambda_0\sin(ks)\;,
\end{equation}
where $k\equiv 2\pi/l_m$ with $l_m$ being the arc length of the ``elementary dipole.'' This model is analytically tractable and the linear response of the extension formally reads:
\begin{eqnarray}
\label{lrPA1}
& &\delta R_{\parallel}=\nonumber\\
& &\frac{E\lambda_0}{k_BT}\int_0^L ds \Big\{\int_0^L ds'\sin(ks')\int_0^{s'}ds''\langle\cos\theta(s)\cos\theta(s'')\rangle_0\nonumber\\
& &-\langle\cos\theta(s)\rangle_0\int_0^L ds' \sin(ks')\int_0^{s'} ds'' \langle\cos\theta(s'')\rangle_0\Big\}\;.
\end{eqnarray}
The explicit final result appears too long and cumbersome and we report it in Appendix B. Here, we only highlight its main features. The sign of the response oscillates depending on the sign of the charge at the tip of the chain (Fig. 3). Specifically, $\delta R_{\parallel}$ will have the same sign as that of the charge in the last polymer segment of arc length $l_m/2$. This is a manifestation of the same phenomenon as the ``odd-even effect'' discussed in Refs. \cite{Schiess2} and \cite{Wink} for free (non-grafted) alternating polyampholytes. In those papers it is shown that a flexible polyampholyte in an external field stretches when the total number of charges is odd, while it collapses when the total number of charges is even (zero net charge). In the case considered here, because of the different boundary conditions, the polymer deforms in the direction of the external field when the charge at the last segment of length $l_m/2$ is positive and it deforms in the opposite direction when this charge is negative. The remarkable point is the strong dependence of the response on the charge of the last segment despite the very large number of segments.


\begin{figure}
\begin{center}
\label{polyamphfig}
\leavevmode
\hbox{%
\epsfxsize=2.5in
\epsffile{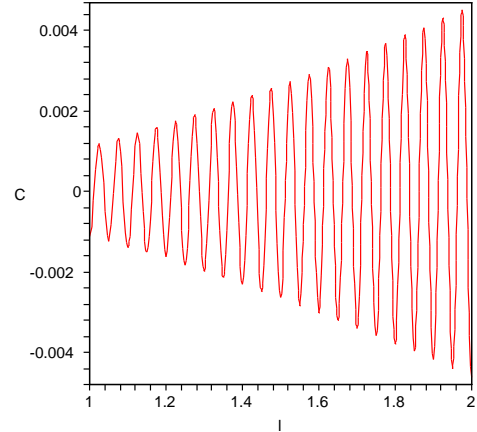}}
\end{center}
\caption{Linear response coefficient for the extension of a periodic polyampholyte, $C\equiv \delta R_{\parallel} k_B T/(E\lambda_0 L_p^3)$, as a function of $l\equiv L/L_p$, for $\theta_0=\pi/2$ and $l_m/L_p=1/20$. }

\end{figure}


In order to understand the interplay between the three characteristic lengths, $L, L_p$, and $l_m$, we focus on the response of a chain with total contour length $L=ml_m$, where m (the number of dipoles) is an integer. In this case, as well as when $L=(m+1/2)l_m$,  the response is peaked. The result reads:
\begin{eqnarray}
\label{lrPA2}
& &\delta R_{\parallel}=\frac{E\lambda_0L_p^3}{k_BT6(9+K^2)(1+K^2)(4+K^2)K^2}\nonumber\\
& &\times\Big\{6K^7-8\pi K^6 m-288\pi m + 2K^7\exp(-2L/L_p)\nonumber\\
& &-8K^7\exp(-L/L_p)+72K^3\exp(-2L/L_p)\nonumber\\
& &-104K^5\exp(-L/L_p)-288K^3\exp(-L/L_p)\nonumber\\
& &+216K^3+26K^5\exp(-2L/L_p)+78K^5-392\pi K^2 m  \nonumber\\
& &-112 \pi K^4 m +\big(\cos^2\theta(0)-\frac{1}{3}\big)\big[78K^5\exp(-2L/L_p)\nonumber\\
& &-297K^3\exp(-L/L_p)-23K^3\exp(-3L/L_p)\nonumber\\
& &-87K^5\exp(-L/L_p)-25K^5\exp(-3L/L_p)\nonumber\\
& &+216K^3\exp(-2L/L_p)+6K^7\exp(-2L/L_p)\nonumber\\
& &-2K^7\exp(-3L/L_p)-6K^7\exp(-L/L_p)\nonumber\\
& &+2K^7+34K^5+104K^3\big]\Big\}\;,
\end{eqnarray}
where $K\equiv 2 \pi L_p/ l_m$.

In the flexible limit ($L\gg L_p$), we obtain:
\begin{eqnarray}
\label{lrPA3}
& &\delta R_{\parallel}=\frac{E\lambda_0L_p^3}{k_BT}\Big\{-\frac{4\pi m}{3 K^2}+\frac{K}{3(1+K^2)(9+K^2)}\nonumber\\
& &\times\big[27+3K^2+\big(\cos^2\theta(0)-\frac{1}{3}\big)(13+K^2)\big]\nonumber\\
& &-O\Big(\exp\big(-\frac{L}{L_p}\big)\Big) \Big\}\;.
\end{eqnarray}
The leading term is the response of a neutral Gaussian chain with a point charge at its free end equal to the {\it average} charge in the last segment of arc length $l_m/2$ of the polyampholyte, which is $-(1/2)\lambda_0l_m / \pi$. The correction from the charge in the bulk depends on the ratio $l_m/L_p$ and it is maximal when $l_m/L_p \approx 2\pi$. We have to distinguish between the case where the chain is flexible on the scale of the elementary dipole, that is, $L_p\ll l_m$, and the case where the chain is weakly bending on that scale. To leading order in $L/L_p$, the relative difference between elongational and bending response defined as  $[\delta R_{\parallel}(\theta_0=0)- \delta R_{\parallel}(\theta_0=\pi/2)]/\delta R_{\parallel}(\theta_0=0)$ is $(2/3)L_p/L$ for $L_p\gg l_m$, and $\sim (L_p/l_m)^2(L_p/L)$ for $L_p\ll l_m$. We point out the qualitative difference between the two cases which could be used as a way to experimentally probe the polyampholyte stiffnes on the scale of $l_m$.

%
%

In the weakly bending limit ($L\ll L_p$), for $\theta(0)\neq(0, \pi)$,
\begin{eqnarray}
\label{lrPA4}
& &\delta R_{\parallel}=-\frac{E\lambda_0L_p^3}{k_BTK^4}\big\{\frac{8}{3}\pi^3m^3+2\pi m\big\}\sin^2\theta(0)+\nonumber\\
& &O\Big(\big(\frac{1}{K}\big)^5\Big)\;,
\end{eqnarray}
whereas for $\theta(0)=(0, \pi)$,
\begin{eqnarray}
\label{lrPA5}
\delta R_{\parallel}=-\frac{E\lambda_0L_p^3}{k_BTK^5}\frac{8}{3}\pi^4m^4+ O\Big(\big(\frac{1}{K}\big)^6\Big)\;.
\end{eqnarray}
For large $m$, as in the flexible limit, we recover the response of a neutral worm-like chain with a point charge at its free end equal to the average charge in the last segment of arc length $l_m/2$ of the polyampholyte. Notice that in the case of elongational or compressional fields, we do not need to assume  large $m$ in order to get this cancellation of the contribution from the charges in the bulk of the chain. 

If we modify the linear charge density of Eq. (\ref{chd}) by introducing a phase, that is, $\lambda(s)=\lambda_0 \sin(ks+\phi)$, the results presented in this Section will remain unchanged up to a phase shift. For example, if we have $\phi=\pi/2$, the response will be peaked when the total contour length is $L=(m+1/4)l_m$ or $(m+3/4)l_m$, where $m$ is an integer.

\section{conclusions}

In this paper we have calculated analytically the linear response of a grafted worm-like chain to a uniform force field. The response assumes scaling forms in the limiting cases of weakly bending and flexible polymers and we have obtained explicit results for polymers of arbitrary stiffness. We have discussed how the response to a field differs from the response to a force exerted at the free end. We have considered a uniformly charged chain and a periodic alternating polyampholyte and we have demonstrated the strong dependence of the response on the distribution of charge along the polymer contour. In the latter case, we have shown how the interplay between the internal persistence length and the characteristic length of the charge modulation affects the response.

\appendix

\section{}

For the sake of completeness and in order to facilitate comparisons, in this Appendix, we cite the main results concerning the linear response of a grafted worm-like chain to a force applied at its free end \cite{KF}. In place of Eq. (\ref{WLCHamint}), the interaction with a force ${\bf F}$ is expressed by adding to ${\cal H}_0$ the term
\begin{equation}
\label{Fint}
{\cal H}_F=-{\bf F}\cdot\int_0^Lds{\bf t}(s)\;.
\end{equation}
Using the orientational correlations of the free chain, one obtains the linear response of the tip orientation,
\begin{eqnarray}
\label{Fres1}
& &\delta\langle\cos\theta(L)\rangle=\frac{FL_p}{k_BT}\Big\{\frac{1}{3}-\frac{2}{3}\exp(-L/L_p)+\frac{1}{3}\exp(-2L/L_p)\nonumber\\
& &+\big(\cos^2\theta(0)-\frac{1}{3}\big)\big[\exp(-2L/L_p)\nonumber\\
& &-\frac{1}{2}\exp(-L/L_p)-\frac{1}{2}\exp(-3L/L_p)\big]\Big\}\;,
\end{eqnarray}
and the linear response of the extension,
\begin{eqnarray}
\label{Fres2}
& &\delta R_{\parallel}=\frac{FL_p^2}{k_BT}\Big\{-1+\frac{2}{3}\frac{L}{L_p}+\frac{4}{3}\exp(-L/L_p)\nonumber\\
& &-\frac{1}{3}\exp(-2L/L_p)+\big(\cos^2\theta(0)-\frac{1}{3}\big)\big[-\frac{1}{3}\nonumber\\
& &+\frac{1}{3}\exp(-3L/L_p)+\exp(-L/L_p)\nonumber\\
& &-\frac{1}{3}\exp(-2L/L_P)-\exp(-2L/L_p)\big]\Big\}\;.
\end{eqnarray}

\onecolumngrid

\section{}

In this Appendix, we report the explicit result for the linear response of a periodic alternating polyampholyte of arbitrary length as obtained from Eq. (\ref{lrPA1}).
\begin{eqnarray}
& &\delta R_{\parallel}=-\frac{E\lambda_0 L_p^3}{k_B T 6(9+K^2)(1+K^2)(4+K^2)K^2}\Big\{108K^5\exp(-l)\cos(Kl)+8K^7\exp(-l)\cos(Kl)\nonumber\\
& &+340K^3\exp(-l)\cos(Kl)-144\sin(Kl)-2K^6\exp(-2l)\sin(Kl)-26K^4\exp(-2l)\sin(Kl)\nonumber\\
& &-2K^7\exp(-2l)\cos(Kl)+4K^5-52K^3\exp(-l)-30\sin(Kl)K^4-144K\exp(-l)-4K^5\exp(-l)\nonumber\\
& &+144K+4\cos(Kl)K^7l+56\cos(Kl)k^5l+144\cos(Kl)Kl-26K^5\exp(-2l)\cos(Kl)+196\cos(Kl)K^3l\nonumber\\
& &+144\exp(-l)\cos(Kl)K-124K^2\sin(Kl)-2K^6\sin(Kl)-6\cos(Kl)K^7-268\cos(Kl)K^3-82\cos(Kl)K^5\nonumber\\
& &+52K^3+52\exp(-l)\sin(Kl)K^4-144\cos(Kl)K+144\exp(-l)\sin(Kl)K^2-72K^3\exp(-2l)\cos(Kl)\nonumber\\
& &-72K^2\exp(-2l)\sin(Kl)+4K^6\exp(-l)\sin(Kl)+\big(\cos^2 \theta(0)-\frac{1}{3}\big)\big[81K^5\exp(-l)\cos(Kl)\nonumber\\
&  &+255\exp(-l)K^3\cos(Kl)+6\exp(-l)K^7\cos(Kl)+23\exp(-3l)K^3\cos(Kl)+108\cos(Kl)\exp(-l)K\nonumber\\
& &+25\exp(-3l)\cos(Kl)K^5+45\exp(-3l)K^4\sin(Kl)+2\exp(-3l)K^7\cos(Kl)+3\exp(-3l)K^6\sin(Kl)\nonumber\\
& &+42\exp(-3l)K^2\sin(Kl)-216K^3\exp(-2l)-78K^5\exp(-2l)\cos(Kl)-6K^6\exp(-2l)\sin(Kl)\nonumber\\
& &-6K^7\exp(-2l)\cos(Kl)+6\exp(-l)K^5+42\exp(-l)K^3-108\exp(-l)K+39K^4\exp(-l)\sin(Kl)\nonumber\\
& &-98\cos(Kl)K^3-28\cos(Kl)K^5+108\exp(-l)K^2\sin(Kl)+3\exp(-l)K^6\sin(Kl)-72\cos(Kl)K\nonumber\\
& &+72K-6K^5+72K-6K^5+6K^3-2\cos(Kl)K^7-216K^2\exp(-2l)\sin(Kl)\nonumber\\
& &-78K^4\exp(-2l)\sin(Kl)\big]\Big\}\;,
\end{eqnarray}
where $l\equiv L/L_p$.

\twocolumngrid

\end{document}